# TPM2.0-Supported Runtime Customizable TEE on FPGA-SoC with User-Controllable vTPM

Jingkai Mao, Xiaolin Chang

## ABSTRACT

Constructing a Trusted Execution Environment (TEE) on Field Programmable Gate Array System on Chip (FPGA-SoC) in Cloud can effectively protect users' private intellectual Property (IP) cores. In order to facilitate the widespread deployment of FPGA-SoC TEE, this paper proposes an approach for constructing a TPM 2.0-compatible runtime customizable TEE on FPGA-SoC. This approach leverages a user-controllable virtual Trusted Platform Module (vTPM) that integrates sensitive operations specific to FPGA-SoC TEE. It provides TPM 2.0 support for a customizable FPGA-SoC TEE to dynamically measure, deploy, and invoke IP during runtime. Our main contributions include: (i) Propose an FPGA-vTPM architecture that enables the TPM 2.0 specification support for FPGA-SoC TEE; (ii) Explore the utilization of FPGA-vTPM to dynamically measure, deploy, and invoke users' IPs on FPGA-SoC TEE; (iii) Extend the TPM command set to accommodate the sensitive operations of FPGA-SoC TEE, enabling users to perform sensitive tasks in a secure and verifiable manner according to the TPM 2.0 specification. We implement a prototype of TRCTEE on the Xilinx Zynq UltraScale+ MPSoC platform and conducted security analysis and performance evaluations to prove the practicality and enhanced security features of this approach.

## 1. Introduction

Cloud service providers are now introducing hardware-based acceleration devices such as Field Programmable Gate Array (**FPGA**) [1], graphics processing unit (GPU) [2], and Application Specific Integrated Circuit (ASIC) [3] to meet the demand for massively parallel computing. In particular, FPGA is widely used by CSPs such as Amazon, Microsoft, and Alibaba, for their dynamic reconfigurability with higher flexibility and energy efficiency [4]. However, the security issues of the cloud data centers [5] decrease end users' confidence in cloud FPGAs and then prevent end users from moving sensitive work such as users' Intellectual Properties (IPs) to clouds. Major CSPs are also exploring how to address the issues of data security and IP protection [6].

Trusted Execution Environment (**TEE**) [7] is a security solution, which can provide a secure and verifiable execution environment for sensitive data and code that is physically isolated from insecure runtime environments. There existed researches to explore TEE support on FPGA devices [8]-[11], especially for building FPGA-System on Chip (**FPGA-SoC**) TEE [12]-[18]. This is because FPGA-SoCs are significantly more secure than pure FPGA architectures because they add processing units containing built-in ARM TrustZone technology to FPGAs. It enables the construction of hardware-isolated FPGA-SoC TEEs. However, existing FPGA-SoC TEE solutions lack of dynamic measurement and verification of sensitive operations for customizing TEE at its runtime, downgrading the user trust. Sensitive operations considered in this paper include dynamically deploying and invoking the user's IPs. Moreover, the lack of unified security standards has also limited their widespread adoption on the cloud [19].

The Trusted Platform Module 2.0 (**TPM 2.0**) standards [20] in Trusted Computing (TC) technology as a unified security standard are widely used in cloud TEE [21]-[25]. Therefore, utilizing TPM to unify the security standards of FPGA-SoC TEEs and provide support for their sensitive operations to build FPGA-SoC TEEs based on TPM standards can effectively solve the above problems.

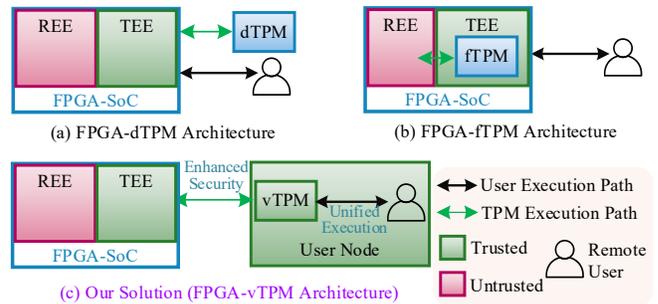

**Figure 1: State-of-the-art Solutions and Our Solution**

The common existing TPM implementations compliant with the TPM 2.0 specification include three main types: (i) firmware TPM (fTPM) [26] deployed in firmware, (ii) virtual TPM (**vTPM**) [27] as a software-only implementation of the TPM, and (iii) discrete TPM (dTPM) as a standalone chip[28]-[31]. It is noticed that the dTPM type approaches can support pure FPGAs [28]-[30], but none of them can be applied to FPGA-SoCs. There exists researches to provide TPM support for FPGA-SoC TEEs, which can be categorized into the following types according to the type of TPM:

**(Solution a) FPGA-dTPM Architecture.** The architecture for using dTPM to provide TC support for FPGA-SoC TEE for secure boot proposed in [31] is shown in Figure 1 (a). This architecture can enhance the FPGA-SoC's security in the boot process with dTPM.

**(Solution b) FPGA-fTPM Architecture**. Implementing fTPM into a TEE built by TrustZone in the FPGA-SoC [13] is illustrated in Figure 1 (b). This architecture can protect fTPM implementation with TEE and improve FPGA-SoC security with fTPM.

Although these existing TPM-enabled FPGA-SoC TEE solutions enhance the security of FPGAs, there are still some problems. For example, **Solution a** uses dTPM which has many security issues in clouds [32] that make it difficult to be applied on cloud FPGA-SoC TEEs. **Solution b** solves the problems of **Solution a**, but still has some problems such as unreliable fTPM that degrades the user's confidence in using it, as detailed in Section 4.2.

All the above discussions motivate our work. This paper explores an approach (denoted as **TRCTEE**, *T*PM2.0-Compatible *R*untime *C*ustomizable *TEE* on FPGA-SoC). TRCTEE can build a runtime customizable TEE on FPGA-SoCs with user-controllable vTPM to provide users with not only a security-enhanced TEE but also a TPM 2.0-compatible method for executing sensitive operations. Specifically, our approach proposes the FPGA-vTPM architecture as shown in Figure 1 (c). User-side reliable vTPM provides TPM support for FPGA-SoC TEEs. The users can utilize this vTPM to perform sensitive operations in FPGA-SoC TEEs in a unified manner. Based on this architecture, we further enhance the security of FPGA-SoC TEE by combining the TPM 2.0 standard. Meanwhile, we design a dynamic measurement methodology as well as a unified execution method for sensitive operations on the user's IP, e.g., IP deployment and invocation. Furthermore, TRCTEE leverages vTPM to build TEEs with wide applicability and has the potential to integrate with other vTPM-based TEE architectures to co-construct a unified TEE platform in clouds. We implemented a prototype of TRCTEE on Xilinx Zynq UltraScale+ XCZU15EG 2FFVB1156 MPSoC, analyzed the security capabilities, and evaluated the performance to indicate TRCTEE's practicability.

We summarize the main contributions as follows:

1) We propose a new architecture (FPGA-vTPM) that provides a more secure TPM for FPGA-SoC TEE. We explore TPM-enabled FPGA-SoC architecture using vTPM. Specifically, we implement a user-controllable vTPM at the user side and design secure initialization, secure communication, and dynamic session key update schemes by combining techniques such as Static Random Access Memory (SRAM) Physical Unclonable Function (PUF) and Public Key Infrastructure authentication. This ensures that the vTPM used by the FPGA-SoC TEE is always secure and trusted.

2) We propose a security-enhanced runtime customizable FPGA-SoC TEE based on FPGA-vTPM architecture. We explore the use of TPM to measure and record sensitive operations on user's IPs located in FPGA-SoC TEE. Specifically, we enhance the secure boot scheme for FPGA-SoCs by utilizing the measurement of the vTPM. Moreover, we implement the user's IP deployments and IP invocations in the TEE and utilize vTPM to record these operations for user verification.

3) We extend the set of TPM commands and responses to provide users with a unified method of executing sensitive operations. We extend the TPM operation set to incorporate the sensitive operations of FPGA-SoC TEEs into the TPM 2.0 standard. Specifically, we design commands/responses that conform to the TPM 2.0 standard for three sensitive operations: updating session keys, deploying user's IPs, and invoking user's IPs. Users can execute sensitive operations directly with these commands, thus unifying operations on FPGA-SoC TEEs.

The structure of this paper is organized as follows: Section 2 presents background information. Section 3 outlines the threat model and requirements. Section 4 reviews related work. Sections 5 and Section 6 detail the design and implementation of our approach. In Section 7, we analyze and evaluate the security and performance of TRCTEE. Section 8 concludes the paper.

## 2. Background

This section presents the background, including the ZYNQ Ultrascale+ Platform and built-in security technology in Section 2.1 and the TPM specification in Section 2.2.

### 2.1 ZYNQ Ultrascale+ Platform and Built-in Security Technology

The Zynq Ultrascale+ MPSOC series devices, introduced by Xilinx, represent a heterogeneous multiprocessor scalable platform and include ARM TrustZone technology as a built-in security technology.

**ZYNQ Ultrascale+ Platform.** This device integrates a high-performance Processing System (PS) and PL [1]. The PL includes hardware-programmable FPGA structures and configuration memory. The hardware logic functions of the FPGA are determined by the bitstream stored in the configuration memory. Configuring the bitstream involves programming a bitstream, which is composed of multiple IP cores, into the configuration memory. The PS comprises the ARM Cortex-A53 Application Processing Unit (APU), the Platform Management Unit (PMU), and the Configuration Security Unit (CSU). The CSU integrates cryptographic primitives and key storage modules necessary for trusted boot mechanisms and implements anti-tamper detection and response mechanisms. Additionally, the CSU includes the Processor Configuration Access Port (PCAP) for bitstream configuration and built-in Direct Memory Access (DMA), referred to as CSUDMA. The PMU firmware (PMU_FW) is primarily responsible for power management and monitoring of system components. The device also includes two types of system memory: On-Chip Memory (OCM) and Double Data Rate (DDR) external memory [33]. The OCM is a type of SRAM used for storing sensitive data and code, such as the First Stage Boot Loader (FSBL) [34] during boot processes.

**ARM TrustZone**. ARM TrustZone is a technology that creates isolation between the TEE and the Rich Execution Environment (REE) within a processor [35]. ARM Trusted Firmware (ATF) serves as the reference implementation for the secure monitor, managing state transitions, and communication between TEE and REE. OP-TEE [36], an open-source TEE software framework, supports running Linux as the REE operating system (OS) and OP-TEE as the TEE OS on the APU. OP-TEE provides the APIs [37][38] for developing Trusted Applications (TAs) and Client Applications (CAs). Each TA is uniquely identified by a Universally Unique Identifier (UUID). ARM TrustZone also offers a method to extend security to the PL. Setting security bits in the AXI Interconnect IP cores and introducing the AWPROT and ARPROT control signals in the AXI slave devices can prevent unauthorized

REE access to secure IP cores belonging to the TEE [39]. Therefore, the security of the PS TEE can be extended to the PL.

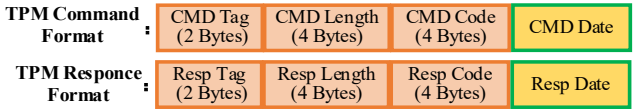

Figure 2: TPM Commands/Responses Format.

## 2.2 TPM Specification

The TPM 2.0 specification [40] is proposed by the Trusted Computing Group (TCG). It supports a broader range of functions, algorithms, and capabilities. Common implementations include the vTPM [27]. It supports the TCG standards of Root of Trust for Storage (RTS) and Root of Trust for Measurement (RTM). The RTM, controlled by the Core Root of Trust for Measurement (CRTM), performs integrity measurements via hashing. The RTS aggregates these integrity measurements and extends them to the Platform Configuration Registers (PCRs). The vTPM typically supports 24 PCRs ($PCR_0$-$PCR_{23}$).

The TPM standard defines the communication format between TPM and external entities [41], namely TPM commands and responses, as illustrated in Figure 2. A TPM command comprises a fixed-length 10-byte command header (depicted in orange in Figure 2) and a variable-length command body (shown in yellow). The command header includes a 2-byte tag identifying the command type, a 4-byte total command length, and a 4-byte command code. The command body contains the parameters and data necessary for the specific command. Similarly, a TPM response consists of a 10-byte response header and a variable-length response body. The response header includes a 4-byte response code that represents the outcome of the command execution. The response body contains the results and data returned from the execution of the command.

Note that the vTPM instance is implemented as software within the OS and its communication is unprotected. Therefore, specific protection measures are required to ensure the security and trustworthiness of the vTPM.

## 3. Threat Model and Requirements

This section describes the threats to TRCTEE in Section 3.1. Section 3.2 gives the function requirements and the security requirements based on the threats.

### 3.1 Threat Model

To achieve the goal of building a TPM2.0-compatible runtime customizable TEE on FPGA-SoC using user-controllable vTPM, the following threats need to be considered.

**Threats against the vTPM.** The vTPM needs to ensure its security and trustworthiness throughout its lifecycle. TRCTEE uses a user-controllable vTPM running on the user side that can ensure the security of the data and code at rest and in use after proper initialization. However, the security of the transmission cannot be guaranteed. Therefore, the vTPM is exposed to the following main threats:

1) Incorrectly vTPM initialization. An adversary may maliciously initialize the vTPM by constructing malicious data causing the vTPM to enter an abnormal state and to fail to provide secure TC functionality.

2) Transmitted data leakage. An adversary can listen to the communication data between the FPGA-SoC and the vTPM or modify it to obtain the sensitive data.

**Threats against the FPGA-SoC.** Adversaries wishing to gain access to the user's IP and data in the FPGA-SoC [9] are the main threats to the FPGA-SoC as follows:

1) IP and data leakage. An adversary may wish to obtain data about private IPs from static bitstream files or from deploying bitstreams to the PL [15]. As well, an adversary may obtain or tamper with the private IPs of the user deployed to the PL, e.g., by injecting a malicious IP into the PL.

2) Unauthorized access to FPGA-SoC. An adversary can read back bitstreams deployed in the PL by reconfiguration interfaces [15], as well as obtain data by unauthorized invocation to user-deployed IPs, etc.

Note that, physical attacks utilizing hardware are not considered in our threat model. Meanwhile, denial-of-service attacks are also not considered in our threat model.

### 3.2 Requirements

We give the function requirements (**FR**) and the security requirements (**SR**) that TRCTEE needs to fulfill.

**Function Requirements:**

To accomplish the goal, TRCTEE needs to satisfy **FR1-FR3**:

1) **FR1: Establishing vTPM Architecture for FPGA-SoC.** Build FPGA-SoC-enabled vTPM architecture (FPGA-vTPM) that meets existing TPM 2.0 APIs available and supports the full TPM 2.0 standard.

2) **FR2: Creating Runtime Customizable FPGA-SoC TEEs with vTPM Support.** Enhance the security of FPGA-SoC TEE runtime customization, e.g., deployment and invocation IPs, by utilizing the TC capabilities (e.g., measurements) provided by vTPM.

3) **FR3: Extending the TPM Command Set to Support FPGA-SoC TEE.** Extend some of the sensitive operations required for FPGA-SoC TEE to the TPM command set to build a unified FPGA-SoC TEE based on the existing TPM 2.0 standard.

**Security Requirements:**

To against the threat given in the threat model, the TRCTEE needs to fulfill **SR1-SR5**:

1) **SR1: Secure Initialization and Binding of vTPM.** Need to ensure that the vTPM is properly initialized and there is no data leakage. Moreover, the data of FPGA-SoC needs to be prevented from being redirected to a malicious vTPM.

2) **SR2: Secure Transmission.** All data transferred between the vTPM and the FPGA-SoC need to ensure confidentiality and integrity and should not be leaked to third parties.

3) **SR3: Trusted Boot and Authentication for FPGA-SoC.** FPGA-SoC needs to be trusted booted with vTPM support and can be authenticated.

4) **SR4: Secure and Verifiable Customization of FPGA-SoC TEE.** Need to ensure that only authorized parties can customize the FPGA-SoC TEE, e.g. deploy IPs. Meanwhile, the bitstreams of the IPs need to ensure confidentiality and integrity. The customization process needs to be logged for verification.

5) **SR5: Secure and Verifiable IP Invocation.** Need to ensure that only authorized parties can invoke IPs already securely deployed in FPGA-SoC TEEs and all invocations can be logged and verified with the support of vTPM.

## 4. Related Work

This section focuses on FPGA-SoC TEE solutions and the research on building TPM-enabled FPGAs.

### 4.1 FPGA-SoC TEE

Utilizing the ARM TrustZone technology built into the FPGA-SoC to divide TEEs and REEs across PS and PL [15][18] can effectively solve the problems of other solutions that do not use ARM TrustZone technology [14][16][17] or do not build TEEs on FPGA-SoCs [8]-[13]. Therefore, the following parts focus on analyzing the FPGA-SoC TEE that constructed TEE across PS and PL, and Table 1 summarizes the solutions according to the **SR**s defined in Section 3.2.

Khan et al. in [15] proposed a security framework SFW that divides both PS and PL into TEE and REE. Wang et al. in [18] proposed to build secure runtime customizable TEEs on FPGA-SoC. However, [15] does not consider remote attestation of device authenticity and IPs execution as well as protection of input and output data of the execution process. Our previous work [18] proposed the SrcTEE to construct a secure runtime-customizable TEE on FPGA-SoC, addressing the weaknesses in SFW [15]. However, it did not consider the dynamic measurement of runtime customizable TEE, specifically for IP deployment and invocation. This prevented users from verifying the trustworthiness of the customized FPGA-SoC TEE.

**Table 1: Comparison of TRCTEE and Existing Solutions**

| | | SR1 | SR2 | SR3 | SR4 | SR5 | Type of TPM |
|---|---|---|---|---|---|---|---|
| **Without TPM** | [15] | - | - | × | √ | × | - |
| | [18] | - | - | √ | √ | √ | - |
| **With TPM** | [13] | × | × | √ | √ | × | fTPM |
| | [31] | × | × | √ | × | × | dTPM |
| **Ours** | | √ | √ | √ | √ | √ | vTPM |

\* √ support  √̇ partial support  × not support  - not applicable or unknown.

### 4.2 FPGAs Supported by TPM

Solutions that utilize TPMs to provide TC support for FPGAs include solutions that provide TPM support for pure FPGAs [28]-[30] and solutions that provide TPM support for FPGA-SoCs [13][31]. The solutions for building TPM support for pure FPGAs propose to utilize TPMs to secure embedded systems, measure FPGA configurations, and provide TC support for FPGAs. However, none of these solutions can be applied in FPGA-SoC that are more secure compared to pure FPGAs. Meanwhile, none of them support hardware-isolated TEE. The solution for building TPM support for FPGA-SoCs solves some of the above problems. Next, we present TPM-enabled FPGA-SoC solutions in terms of TPM implementation classifications (introduced in Section 1). Table 1 lists the **SR**s satisfied by our approach and existing approaches.

Nicholas et al. in [31] adopted PUF and **dTPM** to design a secure boot scheme for FPGA-SoC, protecting the security of hardware application bitstreams during boot and runtime. However, the lack of detection of OS and bitstream tampering makes this secure boot scheme unable to trusted boot FPGA-SoCs. Moreover, using dTPM for cloud FPGA-SoC support is a challenge. Because the dTPM is the standalone hardware in CSPs' cloud infrastructure, where initialization and communication security cannot be assured.

Gross et al. in [13] proposed constructing TEE using ARM TrustZone technology built into FPGA-SoC and implementing **fTPM** within the TEE. This approach also enhances fTPM security using PUF as a random number generator and proposes an encrypted bitstream deployment scheme using TPM's sealing function. The solution using fTPM solves the challenge of dTPM solutions that are difficult to implement on cloud FPGA-SoCs, but still have some problems.

In contrast, our approach solves the above challenges. Specifically, our approach uses a user-controllable vTPM running on the user side and designs a secure initialization protocol that can solve the challenge of unreliable TPM (detailed in Section 5.2.1). The secure communication protocol and key update protocol in our approach address the challenge of insecure communication (detailed in Sections 5.2.2). Additionally, our secure TEE construction includes trusted boot (detailed in Section 5.3.1), secure and verifiable IP deployment (detailed in Section 5.3.2), and invocation for FPGA-SoC (detailed in Section 5.3.3) that address the challenge of untrusted operation of FPGA-SoCs.

## 5. TRCTEE Approach

In this section, we describe our approach in detail, including the TRCTEE approach overview, FPGA-vTPM architecture, security-enhanced runtime customizable FPGA-SoC TEE with vTPM, and the TPM commands/responses for FPGA-SoC TEE extended by TRCTEE in Section 5.1-Section 5.4, respectively.

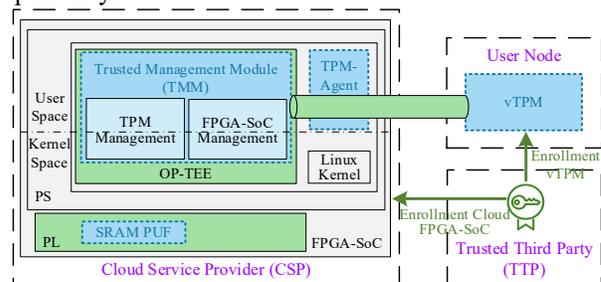

**Figure 3: Illustration of TRCTEE Architecture.**

### 5.1 Overview

The TRCTEE approach consists of an FPGA-vTPM architecture, an approach for constructing the security-enhanced runtime customizable TEEs on FPGA-SoC formed on top of that architecture in conjunction with the FPGA-SoC TEEs proposed in our previous work [18], and an extension of the TPM command/response set applicable to TRCTEE approach.

Here's a brief overview of participants, components, and solutions of the requirements.

**Participants.** As shown in the purple letters in Figure 3, there are three participants:

1) **Trusted Third Party (TTP)**, represents the role of trusted authority for FPGA-SoC device fabrication and initialization in clouds and initialization of user-controllable vTPM.

2) **Cloud Service Provider (CSP)** owns and maintains physical equipment and provides services to remote users over the network.

3) **User Node**, a host owned by the remote user who is the CSP's consumer.

**Components.** As shown in the blue parts in Figure 3, there are four components implemented in TRCTEE (implemented in Section 6).

1) **SRAM PUF** is used to authenticate devices using a challenge-response process utilizing the features of the PUF.

2) **Trusted Management Module (TMM)** is a management module implemented in OP-TEE.

3) **TPM-Agent** is an agent running in REE that forwards encrypted TPM commands and responses.

4) **vTPM** is a vTPM instance running on a user node that supports the full TPM 2.0 standard.

## 5.2 FPGA-vTPM Architecture

This section provides a detailed description of the FPGA-vTPM architecture in the TRCTEE.

### 5.2.1 Initialization

The FPGA-SoC initialization is divided into three steps: device enrollment, device launch, and remote authentication and session key generation.

1) **Device Enrollment.** Device enrollment includes FPGA-SoC for CSPs and vTPM instances for users. For FPGA-SoC, TTP gets and stores the device ID ($\#DI$) generated by the device vendors as the device identification. A customized bootable image is generated for the device, which includes images of FSBL with TTP's public key ($PK_{TTP}$) pre-stored, a full bitstream of SRAM PUF, PMU_FW, ATF, OP-TEE, Linux, and root filesystem (Rootfs) with TMM included. Finally, TTP collects the CRPs of the device.

For vTPM instance enrollment, we default that the owner of the vTPM instance (i.e., the user) is already registered with the TTP, and the TTP can accurately recognize the user's identity and communicate securely with the user. The user then first requests TTP to generate $PK_{TPM}/SK_{TPM}$ of the vTPM instance for this remote authentication process and generate the certificate ($Ca(PK_{TPM})$). After that, TTP stores $Ca(PK_{TPM})$ in the security database for the subsequent authentication process. Finally, TTP sends the $PK_{TPM}/SK_{TPM}$, $Ca(PK_{TPM})$ and $PK_{TTP}$ to user.

2) **Device Launch.** Device launch is also divided into FPGA-SoC launch and vTPM instance launch. For the FPGA-SoC launch, we consider that the bootable image provided by TTP for FPGA-SoC already supports the trusted boot scheme in the TRCTEE approach (detailed in Section 5.3.1).

For the vTPM instance, the user gets the information of the FPGA-SoC device from TTP when launching the vTPM instance.

3) **Remote Authentication and Session Key Generation.** When vTPM launches, it performs remote authentication with the cloud FPGA-SoC and generates a session key ($SessKey$). This protocol combines PUF-based authentication with public key infrastructure, which is divided into 9 steps, as shown in the upper part of the dashed line in Figure 4. See Appendix 2.1 for details on remote authentication and session key generation protocol. Through this protocol, vTPM can be bi-directionally verified with the cloud FPGA-SoC to securely initialize the FPGA-vTPM architecture to support subsequent functionality (Security analysis in Appendix 1).

### 5.2.2 Dynamically Update Session Key

The TRCTEE includes the function to dynamically update the session key. This function allows updating the session key using the present system state without disconnecting the FPGA-SoC from the vTPM. Usually, two situations can trigger the key update process: (i) setting the vTPM's counter threshold and automatically updating the key when the vTPM self-increment counter's value exceeds the set value. (ii) The user sends a key update command ($Update\_CMD$, detailed in Section 5.4.1) to update the key in a user-controllable manner.

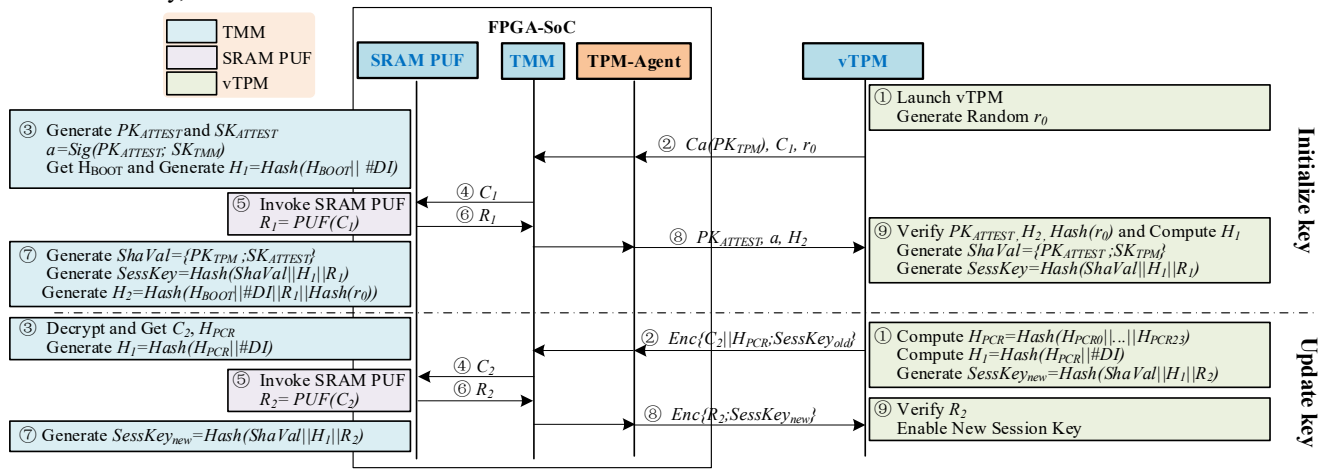

**Figure 4: Key Initialization and Key Update Protocols for TRCTEE.**

Specifically, when a user uses the *Update_CMD* or the counter reaches a set threshold, the vTPM generates a new key based on the current system state (the hash from $PCR_0$ to $PCR_{23}$) in conjunction with a CRP that has never been used. The necessary information is then sent to the TMM for the TMM to update the key. Finally, the two parties verify the new key and use it to communicate. The key update protocol combines public key infrastructure and PUF-based authentication in 9 steps, as shown in the lower part of the dashed line in Figure 4. See Appendix 2.2 for details on dynamically updated session key protocol.

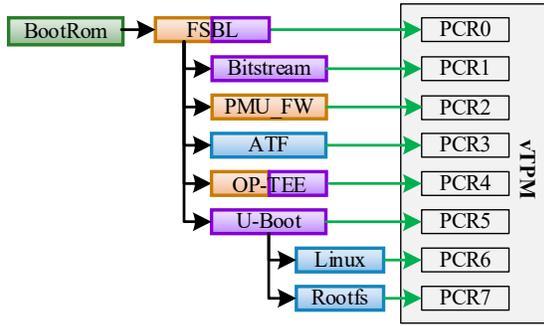

**Figure 5: FPGA-SoC Trusted Chain built by TRCTEE.**

### 5.3 Security-Enhanced Runtime Customizable FPGA-SoC TEE

In this section, we describe the security-enhanced runtime customizable FPGA-SoC TEE constructed by the TRCTEE. It mainly includes a trusted boot scheme for FPGA-SoC and two sensitive operations (user's IPs deployment and IPs invocation) for secure and trusted customized FPGA-SoC TEE at runtime.

#### 5.3.1 Trusted Boot Scheme

The trusted boot scheme refers to the FPGA-SoC secure boot scheme proposed in [18] and combines the TPM 2.0 standards provided by vTPM to form the vTPM-supported FPGA-SoC trusted boot scheme. This scheme uses BootROM as a CRTM to measure and execute securely customized FSBL, the bitstream of SRAM PUF, PMU_FW, ATF, OP-TEE, U-Boot, Linux, and Rootfs, respectively, and extends them to $PCR_0$ to $PCR_7$ of the vTPM, as shown in Figure 5.

#### 5.3.2 User's IP Deployment

Deploying the user's IP is usually realized by invoking PCAP in the CSU or Internal Configuration Access Port (ICAP) in the static logic of the PL. However, in FPGAs divided between TEE and REE, only processes or applications with privileges in the TEE should be allowed to invoke PCAP or ICAP to ensure the trustworthiness of the IPs deployed in the PL and to restrict the attacks [13][15]. In TRCTEE, only the TMM has the privilege to deploy IPs and record the IP deployment operation in the PCR. TRCTEE restricts REE access to PCAP using customized PMU_FW that removes the ability to interact with PCAP and integrates the PCAP and permission checking in the OP-TEE kernel space to make the IP deployment functionality available only to the TMM. Meanwhile, TRCTEE extends the TPM commands (*Deploy_CMD*, detailed in Section 5.4.2) so that the user can only deploy IPs through vTPM and record it, which also ensures that any deploying operation is logged in PCRs for later verification by the user.

When a user wants to deploy IPs, it first creates a bitstream locally or fetches it and encrypts it. The prepared data (detailed in **Assumption 1** in Appendix 2.3) is then stored locally and the encrypted bitstream is uploaded to the file system of the FPGA-SoC. Afterward, the user invokes vTPM using the *Deploy_CMD*. vTPM will collaborate with the TMM to deploy the IPs, measure the deployment process, and extend it to $PCR_8$. The protocol for deploying a user's IP by TRCTEE, which consists of 9 steps, is shown in the upper part of the dashed line in Figure 6. See Appendix 2.3 for details on the user's IP deployment protocol. User's IP Invocation

As with IP deployment, invocation of the IPs should also be isolated from the REE, blocking access from untrustworthy processes and allowing access only from authorized parties in the TEE. In TRCTEE, only the TMM has the invocation privilege and each invocation is logged in the vTPM with the TPM commands (*Invoke_CMD*, detailed in Section 5.4.3) extended by TRCTEE. Specifically, we refer to the scheme in [18] and design an address-based input/output format to invoke the IP and extend the input data to $PCR_9$ and the output data to $PCR_{10}$ to record each invocation. Next, we introduce the IP invocation protocol for TRCTEE, which consists of 9 steps, as shown in the lower part of the dashed line in Figure 6. See Appendix 2.4 for details on the user's IP invocation protocol.

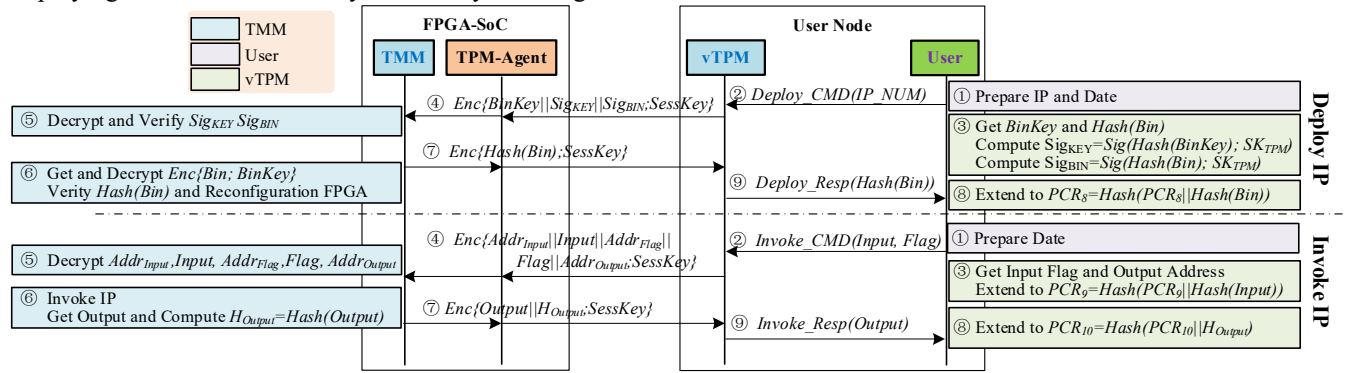

**Figure 6: User's IP Deployment and Invocation Protocols for TRCTEE.**

### 5.4 Extended TPM Commands and Responses

In this section, we introduce the TRCTEE-extended TPM command/response set that enables users to perform sensitive operations on FPGA-SoC TEEs in a unified TPM 2.0 standard, consisting of the command/response for: (i) dynamically updating the key (*Update_CMD/Resp*); (ii) deploying the IP (*Deploy_CMD /Resp*) and (iii) invoking IP (*Invoke_CMD/Resp*).

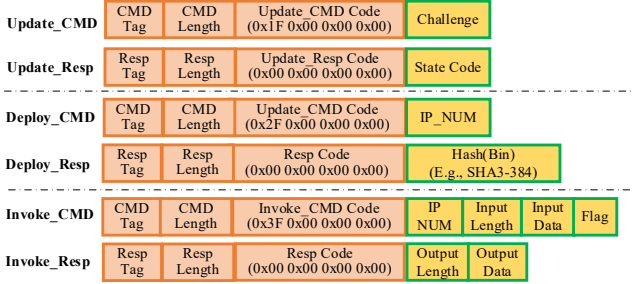

**Figure 7: TRCTEE Extended TPM Command/Response Structure.**

#### 5.4.1 Update_CMD/Resp

*Update_CMD*, a TPM command that is extended to update the session key (detailed in Section 5.2.2). The command length is fixed at 14 bytes, which includes the command header and additional data, as shown in Figure 7. The 10-byte command header includes a 2-byte command tag, a 4-byte command length, and a 4-byte unique command code (0x1F 0x00 0x00 0x00). The additional data is a 4-byte challenge for updating the key.

*Update_Resp* is the response to *Update_CMD*. The length of this response is fixed at 12 bytes, including a 10-byte response header and 2 bytes of additional data. The response header is the same as the response header definition of other responses in the TPM 2.0 standard (detailed in Section 2.2). 2 bytes of additional data is the return code of the execution status of *Update_CMD*. If the return code is 0, it means the session key was successfully updated, and 1 means the update process failed.

#### 5.4.2 Deploy_CMD/Resp

*Deploy_CMD*, which is the TPM command used to deploy a user's IP (detailed in Section 5.3.2). *Deploy_CMD* has a fixed length of 12 bytes, with 10 bytes for the command header containing the special command code (0x2F 0x00 0x00 0x00), and 2 bytes for the IP serial number (*IP_NUM*) of the IP the user wishes to deploy, as shown in Figure 7. The *IP_NUM* with a length of 2 bytes means that it can support up to 65536 IP deployments.

*Deploy_Resp* is the response to *Deploy_CMD*. The response length is fixed at 58 bytes, containing a 10-byte response header and a 48-byte $Hash(Bin)$. $Hash(Bin)$ is the hash of the deployed IP, here the SHA3-384 algorithm is used to compute a 48-byte hash. When successfully deployed, the response code is 0 and the user can verify the hash, otherwise, the response code is 1, representing a deployment failure.

#### 5.4.3 Invoke_CMD/Resp

*Invoke_CMD* is a TRCTEE extended TPM command to invoke the user's IP (detailed in Section 5.3.3). The length of this command is variable and depends on the length of the input data when the user invokes IP, as shown in Figure 7. We define the command format as a 10-byte command header containing a special command code (0x3F 0x00 0x00 0x00) and additional data of variable length. The additional data contains a 2-byte serial number (*IP_NUM*), 4-byte input data length (*Input Length*), input data, and a 4-byte execution state identifier (*Flag*). The *IP_NUM* is used to indicate the IP of this invocation, *Input Length* indicates the number of bytes of the input data. The 2-byte *IP_NUM* can support up to 65536 IP invocations, and the 4-byte *Input Length* supports up to $2^{32}$ bytes of input. In general, the long length of input data should be avoided because the excessively long data will cause the command transmission to be fragmented.

*Invoke_Resp* is the response to the corresponding command, including a 10-byte response header and response data of variable length. The length of the response data depends on the return data of this IP invocation which includes a 4-byte output length (*Output Length*) and output data of variable length. As with commands, the 4-byte output length can support output data of $2^{32}$-byte length, but the output data should not be excessively long to avoid fragmentation.

## 6. Implementation

We implemented a prototype system of our concept. This system consists of an FPGA-SoC device and a vTPM instance running on Linux Server as a user-controllable vTPM. A Xilinx Zynq UltraScale+ XCZU15EG 2FFVB1156 MPSoC is chosen as the FPGA-SoC device, and an AMD server running Ubuntu 22.04 LTS is used as the Linux Server to simulate the user's local environment.

The software stack on the FPGA-SoC follows Xilinx recommendations. We generated the corresponding boot components using the included Xilinx Vitis Design Suite 2020.1, Xilinx Vivado Design Suite 2020.1, and corresponding PetaLinux Tools, and have customized them to support trusted booting of FPGA-SoC that meet our system.

## 7. Evaluation of TRCTEE

This section first analyzes the security of TRCTEE in Section 7.1. Then we evaluate its performance in Section 7.2.

### 7.1 Security Analysis

We will present some of the attacks related to security requirements defined in Section 3.2 (**SR1**-**SR5**), focusing on how TRCTEE can defend against these attacks and thus satisfy **SR**s. Due to space limitations, the detailed formal analysis is given in Appendix.

### 7.2 Performance Evaluation

We evaluated TRCTEE's performance from three aspects: hardware resource utilization, TPM command/response execution time, and the time overhead of the sensitive operation.

Two devices were utilized: one as a user node running a user-controllable vTPM, and the other as an FPGA-SoC device. The user node was a server with an AMD EPYC 7763 64-core processor and 256GB of memory, running Ubuntu 22.04 LTS with kernel version 6.1.0, and a software stack including OpenSSL 1.1.1q and Libtpms v0.9.6. The FPGA-SoC device was a Xilinx Zynq UltraScale+ XCZU15EG 2FFVB1156 MPSoC, running Xilinx-recommended Linux v5.4.0 as the REE OS, and OP-TEE v3.16.0 as the TEE OS. To evaluate TRCTEE's performance, we designed and exported three Convolutional Neural Network (CNN) accelerators [48] using Vivado, with sizes of 2172KB, 672KB, and 372KB respectively.

Table 2: The Reconfigurable Resource Utilization Rate

| Resource | Used | Available | Utilization Rate |
|---|---|---|---|
| LUT | 716 | 341280 | 0.21% |
| LUTRAM | 85 | 184320 | 0.05% |
| FF | 1119 | 682560 | 0.16% |
| CARRY8 | 32 | 42660 | 0.07% |
| F7 Muxes | 37 | 170640 | 0.21% |
| CLB | 255 | 42660 | 0.59% |
| BRAM | 0 | 744 | 0% |

Table 3: Comparison of Resource Consumption

| Ref. | Device | LUT | FF | CLB | BRAM |
|---|---|---|---|---|---|
| 2021[9] | Xilinx Virtex-6 | 78244 | 48048 | - | 14 |
| 2021[10] | Xilinx VCU118 | 21393 | 112922 | - | 30 |
| 2021[11] | Xilinx Virtex-7 | 175854 | 91549 | - | 339 |
| 2019[14] | Xilinx ZCU102 | 48572 | 25719 | - | 84 |
| 2022[16] | AWS F1 | 15573 | - | - | 2 |
| 2021[17] | Xilinx 7000 | 45064 | 44238 | - | 90 |
| 2024[18] | Xilinx AXU15EG | 53989 | 801 | 0 | 0 |
| Ours | Xilinx AXU15EG | 716 | 1119 | 255 | 0 |

\* - indicates that no corresponding reconfigurable resource consumption is published for the existing solutions.

### 7.2.1 Hardware Resource Utilization

Since the implementation of the SRAM PUF components static logic occupies reconfigurable resources in the PL, it is essential to evaluate the reconfigurable resource usage of TRCTEE. We examine the utilization of common reconfigurable resources, including LUT, LUTRAM, FF, CARRY8, F7 Muxes, CLB, and BRAM, as shown in Table 2. We introduce a resource utilization rate to assess the reconfigurable resource usage of TRCTEE.

### 7.2.2 Performance of TPM Commands

TRCTEE provides comprehensive support for the TPM 2.0 standard on FPGA-SoC, and we evaluate its performance using common TPM commands. Due to the secure communication protocol designed within TRCTEE, it is necessary to evaluate the overhead introduced by secure communication. We utilize the tpm2-tools [49] utility in the user space to invoke vTPM, and compare the transmission bytes (**Bytes**) and execution time (**Time**) between TRCTEE and architecture without secure communication:

(i) **Without Enc**, denotes the removal of secure communication protocols in our FPGA-vTPM architecture.

(ii) **Ours**, denotes our approach (TRCTEE).

Table 4 presents the transmission bytes and execution time of common commands.

Table 4: Time for TPM Commands Execution

| tpm2-tools Cmds | TPM Cmds | Without Enc | | Ours | |
|---|---|---|---|---|---|
| | | Bytes | Time | Bytes | Time |
| tpm2_getrandom (16B) | 6 | 481 | 46.7ms | 637 | 109.5ms |
| tpm2_pcrread (SHA384) | 8 | 1595 | 49.4ms | 1803 | 169.4ms |
| tpm2_pcrextend (SHA384) | 4 | 132 | 38.7ms | 236 | 82.0ms |
| tpm2_hash (SHA256) | 4 | 178 | 38.2ms | 282 | 82.2ms |
| tpm2_rsaencrypt (RSA2048) | 24 | 8256 | 53.1ms | 8880 | 360.3ms |
| tpm2_rsadecrypt (RSA2048) | 54 | 14740 | 78.0ms | 16144 | 664.1ms |
| tpm2_encrypt (AES128) | 54 | 7743 | 71.4ms | 9147 | 682.9ms |
| tpm2_decrypt (AES128) | 54 | 7743 | 70.9ms | 9147 | 682.4ms |

## 8. Conclusions and Future Work

This paper presents a TPM2.0-compatible runtime customizable TEE (TRCTEE) on FPGA-SoC, leveraging a user-controllable vTPM to build a runtime customizable TEE on FPGA-SoCs. Our approach integrates the TPM 2.0 specification with FPGA-SoC to enhance security and create a uniform platform for executing sensitive operations.